\documentclass[preprint,aps,12pt,showpacs,nofootinbib,tightenlines]{revtex4}
\usepackage{mathrsfs}
\usepackage{amsmath}
\usepackage{amssymb}
\usepackage{epsfig}
\usepackage{graphicx}
\newcommand{\psl}{ P \hspace{-2.4truemm}/ }

\newcommand{\esl}{ \epsilon \hspace{-1.5truemm}/ }

\def\be{\begin{eqnarray}}
\def\en{\end{eqnarray}}
\def\non{\nonumber\\}

\def\ra{\rangle}

\def\sl{\!\!\!\slash}
\def\prd{{Phys. Rev. D}~}

\def\plb{{ Phys. Lett. B}~}

\def\epjc{{ Eur. Phys. J. C}~}

\begin{document}
%%--------------------------------------------
\title{Insight into $f_0(980)$ through the $B_{(s)}$ charmed decays}
\author{Zhi-Qing Zhang
\footnote{Electronic address: zhangzhiqing@haut.edu.cn}, Si-Yang
Wang, Xing-Ke Ma } %%
\affiliation{\it \small  Department of Physics, Henan University of
Technology, Zhengzhou, Henan 450052, China } %%
\date{\today}
\begin{abstract}

Through analyzing the $B_{(s)}$ charmed decays  $B^{0}\to  \bar
D_{0}f_0(980)$ and $B_{s}\to  \bar D_{0}f_0(980)$ within the
framework of the PQCD factorization approach and comparing with the
current data, we find that there are two possible regions for the
$f_0(980)-f_0(500)$ mixing angle $\theta$: one is centered at
$34^\circ\sim38^\circ$ and the other is falls into
$142^\circ\sim154^\circ$. The former can overlap mostly with one of
allowed angle regions extracted from the decay $B^{0}\to \bar
D_{0}f_0(500)$. The branching fractions of $B_s$ decay modes are
less sensitive to the mixing angle compared with those of B decay
modes. Especially, for the decay $B_{s}\to D^0 f_0(980)$, its
branching fraction changes only slightly between
$(1.2\sim1.8)\times10^{-7}$ when the mixing angle $\theta$ runs from
$0^\circ$ to $180^\circ$. All of our results support the picture
that the $f_0(980)$ is dominated by two quark component in the B
decay dynamic mechanism. Furthermore, the $s\bar s$ component is
more important than the $q\bar q=(u\bar u+d\bar d)/\sqrt{2}$
component. This point is different from $f_0(500)/\sigma$. Last but not
least, our picture is not in conflict with the popular four-quark
explanation.
\end{abstract}

\pacs{13.25.Hw, 12.38.Bx, 14.40.Nd}
\vspace{1cm}

\maketitle

%=======================================================================
%                     Introduction
%=======================================================================

\section{Introduction}\label{intro}
Up to now the quark-level substructure of  scalar mesons is still
not well understood. Especially, the slight scalars mesons,
including $f_0(500)(\sigma), f_0(980), K^*_0(800)(\kappa)$ and
$a_0(980)$, which form an SU(3) flavor nonet and are considered as
either two quark states or  tetraquak states (di-quark and
anti-diquark structure) as originally advocated by Jaffe
\cite{jaffe}. Certainly, there are other different SU(3)
scenarios about scalar mesons \cite{minkowski}. If one considers
these light scalar mesons as two quark states, $q\bar{q}$ structure,
there are experiments indicate that the heaviest one $f_0(980)$ and the
lightest one $f_0(500)$ in this SU(3) nonet must have a mixing \be
|f_0(980)\rangle=|s\bar s\rangle\cos\theta+|n\bar
n\rangle\sin\theta, \;\;\;\;\; |f_0(500)\rangle=-|s\bar
s\rangle\sin\theta+|n\bar n\rangle\cos\theta, \label{mix} \en where $|n\bar
n\rangle\equiv(u\bar u+d\bar d)/\sqrt2$. For the mixing
angle $\theta$, there are several different values from experimental
measurements. A mixing angle $\theta=34^\circ\pm6^\circ$ was determined from
the decays $J/\Psi\to f_0\phi, f_0\omega$, and
$31^\circ\pm5^\circ$ or $42^\circ\pm7^\circ$ from the
decays $D_{(s)}\to f_0(980)\pi, f_0(980)K$, while a range
$35^\circ<|\theta|<55^\circ$ was given from the analysis of three body decay
$D_s^+\to\pi^+\pi^+\pi^-$. An analysis of $f_0(980)-f_0(500)$ mixing
by using the light cone QCD sum rules \cite{gokalp}, yielded
$\theta=27^\circ\pm13^\circ$ and $\theta=41^\circ\pm11^\circ$ . The value of $\theta$
$\sim34^\circ$ or $\sim146^\circ$ was obtained in the decays $B_s\to
J/\psi f_0(980), J/\psi\sigma$ \cite{jingwu}. Ochs \cite{ochs} found
$\theta=30^\circ\pm3^\circ$ by averaging over several decay
processes. The authors of Ref.\cite{lmzhang} provided a limit on the
mixing angle $\theta<29^\circ$ at $90\%$ confidence. As we know, the mixing between $f_0(980)-\sigma$
is something like that in $\eta-\eta^\prime$, but with much more uncertainties. In order to explain the $K-\eta^\prime$
puzzle, some complex mixing mechanisms including gluon even $\eta_c$ meson in $\eta-\eta^\prime$ were
also considered \cite{feldmann1,feldmann2,hnli1}. This led people to conjecture that $f_0(980)$
and $f_0(500)$ may not be simple quark-antiquark states, perhaps there exist more complicated structure except
the $f_0(980)-\sigma$ mixing.

Recently, the decays $B_{(s)}\to \bar{D} f_0(500),\bar{D} f_0(980)$
were measured by LHCb collaboration \cite{lhcb1,lhcb2}: \be
\mathcal{B}(B^0\to
\bar{D}^0f_0(500))&=&(11.2\pm0.8\pm0.5\pm2.1\pm0.5)\times10^{-5},\\
\label{1} \mathcal{B}(B^0\to
\bar{D}^0f_0(980))&=&(1.34\pm0.25\pm0.10\pm0.46\pm0.06)\times10^{-5},\\
\label{2}
\mathcal{B}(B^0_s\to\bar{D}^0f_0(980))&=&(1.7\pm1.0\pm0.5\pm0.1)\times10^{-6}.
\en where the first and the second uncertainties are statistical and
experimental systematic errors, respectively, the third one is
from the model-dependent error. We
 see there exist larger statistical error in the $B^0_s$ decay
and the model-dependent error in the fist two $B^0$ decays. By using
these new data, we will try to constrain the mixing angle between
$f_0(980)$ and $\sigma$ through these $B_{(s)}$ decays in the
perturbative QCD (pQCD) approach. There was a work about
constraining the mixing angle through $B^0_s\to J/\Psi f_0(980),
J/\Psi\sigma$ decays \cite{jingwu}, but two different approaches
were used in the same decay channel: the factorizable contribution
and vertex corrections are calculated in the QCD Factorization
(QCDF) approach, while the hard spectator scattering corrections are
calculated in the pQCD approach. So one may suspect its rationality
and reliability in determining the mixing angle between
$f_0(980)-\sigma$. The B meson decays with a D meson involved in the
final states have been studied in pQCD approach, such as $B\to D P,
DV, DA$ \cite{li,zou,zhang1,zhang2}, here $P,V,A$ represent a
pseudoscalar, vector and axial-vector meson, respectively. Most of
the predictions can well explain the experimental data. While an
explicit calculation for the branching ratio of the decay $B^0_s\to
\bar D^0 f_0(980)$ gives
$(3.5^{+1.26+0.56}_{-1.15-0.77})\times10^{-5}$ \cite{kim}, which is
quite different from the present experimental result. So we would
like to systematically study the decays $B_{(s)}\to \bar D f(980)$
in the pQCD approach, including the CKM suppressed decays
$B_{(s)}\to D f_0(980)$. At last, the decays $B_{(s)}\to \bar D^{*}
f(980), D^{*} f_0(980)$ are also considered.

The layout of this paper is as follows. In Sec.\ref{proper}, decay
constants and light-cone distribution amplitudes of the relevant
mesons are introduced. In Sec.\ref{results}, we then analyze these
decay channels using the PQCD approach. The numerical results and
the discussions are given in Sec. \ref{numer}. Conclusions are
presented in the final part.
%=======================================================================
%       Physical properties of $f_0(980)$ and $f_0(1500)$
%=======================================================================

\section{decay constants and distribution amplitudes }\label{proper}

For the wave function of the heavy $B_{(s)}$ meson, we take \be
\Phi_{B_{(s)}}(x,b)= \frac{1}{\sqrt{2N_c}} (\psl_{B_{(s)}}
+m_{B_{(s)}}) \gamma_5 \phi_{B_{(s)}} (x,b). \label{bmeson} \en Here
only the contribution of the first Lorentz structure $\phi_{B_{(s)}}
(x,b)$ is taken into account, since the contribution of the second
Lorentz structure $\bar \phi_{B_{(s)}}$ is numerically small
\cite{cdlu} and can be neglected. For the distribution amplitude
$\phi_{B_{(s)}}(x,b)$ in Eq.(\ref{bmeson}), we adopt the following
model: \be
\phi_{B_{(s)}}(x,b)=N_{B_{(s)}}x^2(1-x)^2\exp\left[-\frac{M^2_{B_{(s)}}x^2}{2\omega^2_b}-\frac{1}{2}(\omega_bb)^2\right],
\en where $\omega_b$ is a free parameter and taken to be
$\omega_b=0.4\pm0.04 (0.5\pm0.05)$ GeV for $B(B_s)$ in numerical
calculations, and $N_B=101.445$ $(N_{B_s}=63.671)$ is the
normalization factor for $\omega_b=0.4$ $(0.5)$. For $B_s$ meson,
the SU(3) breaking effects are taken into consideration.

As for the wave functions of the $D$ meson, we use the form derived
in Ref.\cite{kurimoto} \be
\int\frac{d^4\omega}{(2\pi)^4}e^{ik\cdot\omega}\langle0|\bar c_\beta(0)u_{\gamma}(\omega)|\bar D^0\rangle&=&-\frac{i}{\sqrt{2N_c}}[(\psl_D+m_D)\gamma_5]_{\gamma\beta}\phi_D(x,b),\\
\int\frac{d^4\omega}{(2\pi)^4}e^{ik\cdot\omega}\langle0|\bar
c_\beta(0)u_{\gamma}(\omega)|\bar
D^{*0}\rangle&=&-\frac{i}{\sqrt{2N_c}}[(\psl_{D^*}+m_{D^*})\esl_L]_{\gamma\beta}\phi^L_{D^*}(x,b),
\en where $\esl_L$ is the longitudinal polarization vector. In this
work only the longitudinal polarization component is used. Here we
take the best-fitted form $\phi_D^{(*)}$ from B to charmed meson
decays derived in \cite{li} as \be
\phi_{D}(x,b)=\frac{f_{D}}{2\sqrt{2N_c}}6x(1-x)[1+C_{D}(1-2x)]\exp[\frac{-\omega^2b^2}{2}].
\en For the wave function $\phi_{D_{s}}(x,b)$, it has the similar
expression as $\phi_D(x,b)$ except with different parameters, and
given as follows: $f_D=204.6$ MeV, $f_{D_s}=257.5$ MeV, and
$C_{D_{(s)}}=0.5$ $(0.4)$, $\omega_{D_{(s)}}=0.1$ $(0.2)$
\cite{pdg14}. For the wave function $\phi_{D^*_{(s)}}(x,b)$, we take
the same distribution amplitude with that of the pseudoscalar meson
$D_{(s)}$ because of their small mass difference, except with
different decay constants $f_{D^*}=270$ MeV and $f_{D^*_s}=310$ MeV
\cite{wanggl}.

Since the neutral scalar meson $f_0(980)$ cannot be produced via the
vector current, we have $\langle f_0(p)|\bar q_2\gamma_\mu
q_1|0\ra=0$ (the abbreviation $f_0$ denotes the $f_0(980)$ for
simplicity). Taking the $f_0(980)-\sigma$ mixing into account, the
scalar current $\langle f_0(p)|\bar q_2q_1|0\ra=m_S\bar {f_S}$ can
be written as: \be \langle f_0^n|d\bar d|0\ra=\langle f_0^n|u\bar
u|0\ra=\frac{1}{\sqrt 2}m_{f_0}\tilde f^n_{f_0},\,\,\,\, \langle
f_0^s|s\bar s|0\ra=m_{f_0}\tilde f^s_{f_0}, \en where $f_0^{(n,s)}$
represent for the quark flavor states for $n\bar{n}$ and $s\bar{s}$
components of $f_0$ meson, respectively. As the scalar decay
constants $\tilde f_{f_0}^n$ and $\tilde f_{f_0}^s$ are very
close\cite{CCYscalar}, we can assume $\tilde f_{f_0}^n=\tilde
f_{f_0}^s$ and denote them as $\bar f_{f_0}$ in the following.

The twist-2 and twist-3 LCDAs for the different components of
$f_0(980)$ are defined by: \be \langle f_0(p)|\bar q(z)_l
q(0)_j|0\rangle &=&\frac{1}{\sqrt{2N_c}}\int^1_0dxe^{ixp\cdot
z}\{p\sl\Phi_{f_0}(x)
+m_{f_0}\Phi^S_{f_0}(x)+m_{f_0}(n\sl_+n\sl_--1)\Phi^{T}_{f_0}(x)\}_{jl},\non
\label{LCDA} \en where we assume $f_0^n(p)$ and $f_0^s(p)$ are the
same and denote them as $f_0(p)$, $n_+$ and $n_-$ are light-like
vectors: $n_+=(1,0,0_T),n_-=(0,1,0_T)$. The normalization of the
distribution amplitudes are related to the decay constants: \be
\int^1_0 dx\Phi_{f_0}(x)=\int^1_0
dx\Phi^{T}_{f_0}(x)=0,\,\,\,\,\,\,\,\int^1_0
dx\Phi^{S}_{f_0}(x)=\frac{\bar f_{f_0}}{2\sqrt{2N_c}}. \en The
twist-2 LCDA $\Phi_{f_0}(x)$ can be expanded in terms of Gegenbauer
polynomials as: \be \Phi_{f_0}(x)=\frac{1}{2\sqrt{2N_c}}\bar
f_{f_0}6x(1-x)\left[B_0+\sum_{m=1}B_mC^{3/2}_n(2x-1)\right], \en
with the decay constant $\bar f_{f_0}=0.18\pm0.015$ GeV \cite{defa}.
It is noticed that all the even Gegenbauer momentums vanish due to
the charge conjugation invariance. As for the odd Genbauer
momentums, only the first term is kept and the value of the
coefficient is taken as $B_1=-0.78\pm0.08$ \cite{CCYscalar}. For the
twist-3 LCDA, we also take the first term of the Gegenbauer
expansion, i.e. the asymptotic form, \be
\Phi^S_{f_0}(x)=\frac{1}{2\sqrt{2N_c}}\bar f_{f_0},
\;\;\;\Phi^T_{f_0}(x)=\frac{1}{2\sqrt{2N_c}}\bar f_{f_0}(1-2x). \en
%===========================================================================
%                    Decay amplitudes in PQCD approach
%============================================================================

\section{ the perturbative QCD  calculation} \label{results}
The weak effective Hamiltonian $H_{eff}$ for the charmed $B_{(s)}$
decays $B_{(s)}\to \bar D f_0(980), \bar D^{*} f_0(980)$, is
composed only by the tree operators and given by: \be
H_{eff}=\frac{G_F}{\sqrt{2}}V^*_{cb}V_{uq}[C_1(\mu)O_1(\mu)+C_2(\mu)O_2(\mu)],
\en where the tree operators are writen as: \be O_1=(\bar
c_{\alpha}b_{\beta})_{V-A}(\bar D_{\beta}u_{\alpha})_{V-A},\;\;\;
O_2=(\bar c_{\alpha}b_{\alpha})_{V-A}(\bar
D_{\beta}u_{\alpha})_{V-A}, \en with $D$ represents $d(s)$. These
decays with larger CKM matrix elements, say the $\bar b\to \bar d$
transition, $|V_{cb}V_{ud}|=0.04$ are called CKM allowed decays.
Another kind of decays $B_{(s)}\to D^0 f_0, D^{*0}f_0,
D^{+}_{(s)}f_0,D^{*+}_{(s)}f_0$ with smaller CKM matrix elements (in
case of $b\to d$ transition, $|V_{ub}V_{cd}|=0.00093$) are called
CKM suppressed decays and the corresponding weak effective
Hamiltonian is given as: \be
H_{eff}=\frac{G_F}{\sqrt{2}}V^*_{ub}V_{cq}[C_1(\mu)O_1(\mu)+C_2(\mu)O_2(\mu)].
\en Here we take the decay $B^0\to \bar D^0 f_0$ as an example,
whose leading-order Feynman diagrams are shown in Figure 1.
\begin{figure}[t]
\vspace{-4cm} \centerline{\epsfxsize=18 cm \epsffile{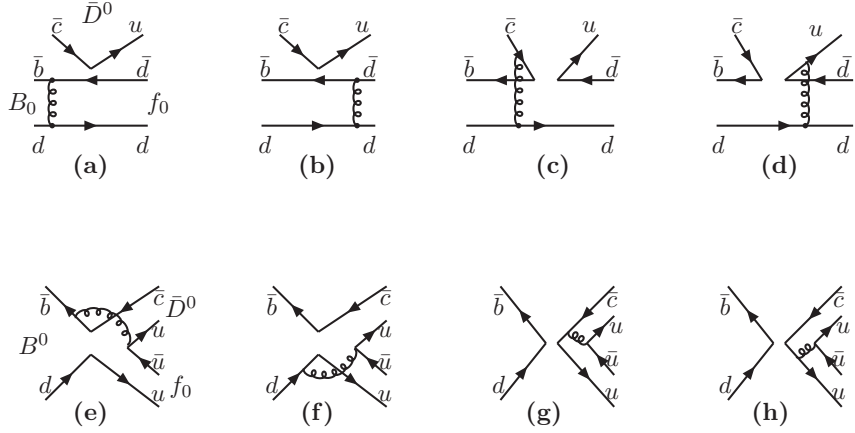}}
\vspace{-15.5cm} \caption{ Diagrams contributing to the $B^0\to \bar
D^0f_0(980)$ decay.}
 \label{Figure1}
\end{figure}
The Feynman diagrams on the first row are for the emission types,
where Figs.(a) and (b) are the factorizable diagrams, Figs.(c) and
(d) are the nonfactorizable ones, their amplitudes are written as:
\be \mathcal{F}^{\bar D}_{B\to f_0}&=&8\pi C_FM^4_Bf_D\int_0^1 d
x_{1} dx_{2}\, \int_{0}^{\infty} b_1 db_1 b_2 db_2\, \phi_B(x_1,b_1)
[(1+x_2)\phi_{f_0}(x_2)+r_f(1-2x_2)\non && \times
(\phi^s_{f_0}(x_2)+\phi^t_{f_0}(x_2))]
E_e(t_a)h_e(x_1,x_2(1-r^2_D),b_1,b_2)S_t(x_2)\non &&
+2r_f\phi_{f_s}(x_2)E_e(t_b)\
h_e(x_2,x_1(1-r^2_D),b_2,b_1)S_t(x_1)],\\
\mathcal{M}^{\bar D}_{B\to f_0}&=&32\pi C_f
m_B^4/\sqrt{2N_C}\int_0^1 d x_{1} dx_{2} dx_{3}\, \int_{0}^{\infty}
b_1 db_1 b_3 db_3\,\phi_B(x_1,b_1)\phi_D(x_3,b_3)\non &&
\times\left\{\left[(x_3-1)\phi_{f_0}(x_2)+r_{f_0}x_2(\phi^s_{f_0}(x_2)-\phi^t_{f_0}(x_2))-4r_{f_0}r_cr_D
\phi^s_{f_0}(x_2)\right] \right.\non &&\left.\times
E_{en}(t_c)h^{c}_{en}(x_1,x_2(1-r_D^2),x_3,b_1,b_3)+E_{en}(t_d)h^{d}_{en}(x_1,x_2(1-r_D^2),x_3,b_1,b_3)
\right.\non && \left. \times\left[(x_2+x_3)\phi_{f_0}(x_2)
-r_{f_0}x_2(\phi^s_{f_0}(x_2)+\phi^t_{f_0}(x_2))\right]\right\}, \en
with the mass ratios $r_{f_0}=m_{f_0}/M_B, r_{D}=m_D/M_B,$ and
$r_{c}=m_c/M_B$. The evolution factors  evolving the scale $t$ and
the hard functions of the hard part of factorization amplitudes are
listed as: \be
E_e(t)&=&\alpha_s(t)\exp[-S_B(t)-S_{f_0}(t)],\label{suda1}\\
E_{en}(t)&=&\alpha_s(t)\exp[-S_B(t)-S_{f_0}(t)-S_D(t)|_{b_1=b_2}],\label{suda2}\\
h_e(x_1,x_2,b_1,b_2)&=&K_0(\sqrt{x_1x_2}m_Bb_1)\left[\theta(b_1-b_2)K_0(\sqrt{x_2}m_Bb_1)
I_0(\sqrt{x_2m_Bb_2})\right.
\non && \left.+\theta(b_2-b_1)K_0(\sqrt{x_2m_Bb_2})I_0(\sqrt{x_2m_Bb_1})\right],\\
h^{j}_{en}(x_1,x_2,x_3,b_1,b_3)&=&\left[\theta(b_1-b_3)K_0(\sqrt{x_1x_2(1-r^2_D)}m_Bb_1)
I_0(\sqrt{x_1x_2(1-r^2_D)}m_Bb_3)\right.\non && \left.+(b_1\leftrightarrow b_3)\right]
\left(\begin{matrix}K_0(A_jm_Bb_3)& \text{for} A^2_j\geq 0\\
\frac{i\pi}{2}H^{(1)}_0(\sqrt{|A^2_j|}m_Bb_3)&\text{for} A^2_j\leq
0\\\end{matrix}\right), \en with the variables $A^2_j(j=c,d)$ listed
as: \be
A^2_c&=&r_c^2-(1-x_1-x_3)(x_2(1-r^2_D)+r_D^2),\\
A^2_d&=&(x_1-x_3)x_2(1-r^2_D). \en The hard scale $t$  and the
expression of Sudakov factor in each amplitude can be found in the
Appendix. As we know, the double logarithms $\alpha_sln^2x$ produced
by the radiative corrections are not small expansion parameters when
the end-point region is important. In order to improve the
perturbative expansion, threshold resummation of these logarithms to
all order is needed, which leads to a quark jet function: \be
S_t(x)=\frac{2^{1+2c}\Gamma(3/2+c)}{\sqrt{\pi}\Gamma(1+c)}[x(1-x)]^c,
\en with $c=0.32$. It is effective to smear the end point
singularity with a momentum fraction $x\to0$. This factor will also
appear in the factorizable annihilation type amplitudes.

The amplitudes for the Feynman diagrams on the second row can be
obtained by the Feynman rules and are given as: \be
\mathcal{M}^{\bar D}_{ann}&=&32\pi C_f m_B^4/\sqrt{2N_C}\int_0^1 d
x_{1} dx_{2} dx_{3}\,\int_{0}^{\infty} b_1 db_1 b_3 db_3\,
\phi_B(x_1,b_1)\phi_D(x_3,b_3)\non &&
\times\left\{E_{an}(t_{e})h^e_{an}(x_1,x_2,x_3,b_1,b_3)\left[x_3\phi_{f_0}(x_2)
\right.\right.\non && \left.\left.
+r_Dr_{f_0}((x_2-x_3-3)\phi^s_{f_0}(x_2)+(x_2+x_3-1)\phi^t_{f_0}(x_2))
\right]\right.\non &&\left.
+E_{an}(t_{f})h^f_{an}(x_1,x_2,x_3,b_1,b_3)\left[(x_2-1)
\phi_{f_0}(x_2) \right.\right.\non &&\left.\left.
+r_Dr_{f_{0}}((1+x_3-x_2)\phi^s_{f_{0}}(x_2)+(x_2+x_3-1)\phi^t_{f_{0}}(x_2))\right]\right\},
\en \be \mathcal{F}^{\bar D}_{ann}&=&-8\pi C_f f_{B}m_B^4\int_0^1 d
x_{2} dx_{3}\, \int_{0}^{\infty} b_2 db_2 b_3 db_3\,
\phi_D(x_3,b_3)\left\{\left[(1-x_2)\phi_{f_{0}}(x_2)-2r_{f_0}r_D\right.\right.\non
&& \left.\left.\times
x_2\phi^t_{f_0}(x_2)+2r_Dr_{f_{0}}(x_2-2)\phi^s_{f_{0}}(x_2)\right]
E_{af}(t_g)h_{af}(x_3,(1-x_2)(1-r_D^2),b_3,b_2)\right.\non &&\left.
+E_{af}(t_h)h_{af}(x_2,x_3(1-r_D^2),b_2,b_3)\left[-x_3\phi_{f_{0}}(x_2)
+2r_Dr_{f_{0}}(x_3+1)\phi^s_{f_{0}}(x_2)\right]\right\}. \en
Similarly, $F^{\bar D}_{ann}(M^{\bar D}_{ann})$ are the
(non)factorizable annihilation type amplitudes, where the evolution
factors $E$ evolving the scale $t$ and the hard functions of the
hard part of factorization amplitudes are listed as: \be
E_{an}(t)&=&\alpha_s(t)\exp[-S_B(t)-S_D(t)-S_{f_0}(t)|_{b_2=b_3}],\label{suda3}\\
E_{af}(t)&=&\alpha_s(t)\exp[-S_D(t)-S_{f_0}(t)],\label{suda4}\\
h_{an}^j(x_1,x_2,x_3,b_1,b_3)&=&i\frac{\pi}{2}\left[\theta(b_1-b_3)H^{(1)}_0(\sqrt{x_2x_3(1-r^2_D)}m_Bb_1)
J_0(\sqrt{x_2x_3(1-r^2_D)}m_Bb_3)\right.\non &&\left.+(b_1\leftrightarrow b_3)\right]
\left(\begin{matrix}K_0(L_jm_Bb_1)& \text{for} L^2_j\geq 0\\
\frac{i\pi}{2}H^{(1)}_0(\sqrt{|L^2_j|}m_Bb_1)& \text{for} L^2_j\leq 0\\ \end{matrix}\right),\\
h_{af}(x_2,x_3,b_2,b_3)&=&(i\frac{\pi}{2})^2H^{(1)}_0(\sqrt{x_2x_3}m_Bb_2)\non &&
\times[\theta(b_2-b_3)H^{(1)}_0(\sqrt{x_3}m_Bb_2)J_0(\sqrt{x_3}m_Bb_3)+(b_2\leftrightarrow b_3)],
\en
where the definitions of $L^2_j(j=e,f)$ are written as:
\be
L^2_e&=&r_b^2-(1-x_3)(1-(1-x_2)(1-r^2_D)-x_1),\\
L^2_f&=&x_3(x_1-(1-x_2)(1-r^2_D)). \en The functions $H^{(1)}_0,J_0,
K_0, I_0$, which appear in the upper hard kernel
$h_{e},h^j_{en},h^j_{an},h_{af}$ are the (modified) Bessel
functions, which are obtained from the Fourier transformations of
the quark and gluon propagators. Combining above amplitudes, one can
easily to write down the total decay amplitudes of each considered
channel \be \mathcal{A}(B^0\to\bar
D^0f_0(980))&=&\frac{G_F}{\sqrt2}V^*_{cb}V_{ud}(\emph{F}^{\bar
D}_{B\to f_0}a_2
+\emph{M}^{\bar D}_{B\to f_0}C_2+{M}^{\bar D}_{ann}C_2+\emph{F}^{\bar D}_{ann}a_2),\\
\mathcal{A}(B^0\to D^0f_0(980))&=&\frac{G_F}{\sqrt2}V^*_{ub}V_{cd}(\emph{F}^{D}_{B\to f_0}a_2
+\emph{M}^{D}_{B\to f_0}C_2+\emph{M}^{f_0}_{ann}C_2+\emph{F}^{f_0}_{ann}a_2),\\
\mathcal{A}(B^0_s\to\bar D^0f_0(980))&=&\frac{G_F}{\sqrt2}V^*_{cb}V_{us}(\emph{F}^{D}_{B\to f_0}a_2
+\emph{M}^{D}_{B\to f_0}C_2+\emph{M}^{D}_{ann}C_2+\emph{F}^{D}_{ann}a_2),\\
\emph{A}(B^0_s\to D^0f_0(980))&=&\frac{G_F}{\sqrt2}V^*_{ub}V_{cs}(\emph{F}^{D}_{B\to f_0}a_2
+\emph{M}^{D}_{B\to f_0}C_2+\emph{M}^{f_0}_{ann}C_2+\emph{F}^{f_0}_{ann}a_2),\\
\mathcal{A}(B^+\to D^+f_0(980))&=&\frac{G_F}{\sqrt2}V^*_{ub}V_{cd}(\emph{F}^{D}_{B\to f_0}a_1
+\emph{M}^{D}_{B\to f_0}C_2/3+\emph{M}^{f_0}_{ann}C_2/3+\emph{F}^{f_0}_{ann}a_1),\\
\mathcal{A}(B^+\to
D^+_sf_0(980))&=&\frac{G_F}{\sqrt2}V^*_{ub}V_{cs}(\emph{F}^{D}_{B\to
f_0}a_1 +\emph{M}^{D}_{B\to
f_0}C_2/3+\emph{M}^{f_0}_{ann}C_2/3+\emph{F}^{f_0}_{ann}a_1), \en
and likewise for the corresponding decays with the pseudoscalar
meson $D$ replaced by the vector meson $D^*$.
%===========================================================================
%                  Numerical results and discussions
%============================================================================
\section{Numerical results and discussions for $B_{(s)}$ Decays} \label{numer}

We use the following input parameters for numerical calculations
\cite{pdg14}: \be
f_B&=&190 MeV, f_{B_s}=230 MeV, M_B=5.28 GeV, M_{B_s}=5.37 GeV, \\
\tau_B^\pm&=&1.638\times 10^{-12} s,\tau_{B^0}=1.519\times 10^{-12} s, \tau_{B_s}=1.512\times 10^{-12} s,\\
M_{D^0}&=&1.869 GeV, M_{D_s^+}=1.968 GeV, M_{D^{*0}}=2.007 GeV,
M_{D_s^{*+}}=2.112 GeV. \en For the CKM matrix elements, we adopt
the Wolfenstein parametrization and the updated values $A=0.814,
\lambda=0.22537, \bar\rho=0.117\pm0.021$ and
$\bar\eta=0.353\pm0.013$ \cite{pdg14}.

In the $B_{(s)}$-rest frame, the decay rates of $B_{(s)}\to D^{(*)}_{(s)}f_0(980)$
can be written as:
\be \mathcal{BR}(B_{(s)}\to D^{(*)}_{(s)}f_0(980))=\frac{\tau_{B_{(s)}}}{16\pi M_B}(1-r^2_{D^{(*)}_{(s)}}){\cal A}, \en
where $\cal A$ is the total decay amplitude of each considered decay, which has been given in last section.

Using the input parameters and the wave functions as specified in
this section and Sec.\ref{proper}, we give the dependencies of the
branching ratios $\mathcal{BR}(B^0\to \bar D^0f_0(980))$ and
$\mathcal{BR}(B_{s}\to \bar D^0f_0(980))$ on the mixing angle
$\theta$ shown in Fig.2.
\begin{figure}[thb]
\begin{center}
\vspace{-0.5cm} \centerline{\hspace{5.0cm}\epsfxsize=15.0cm
\epsffile{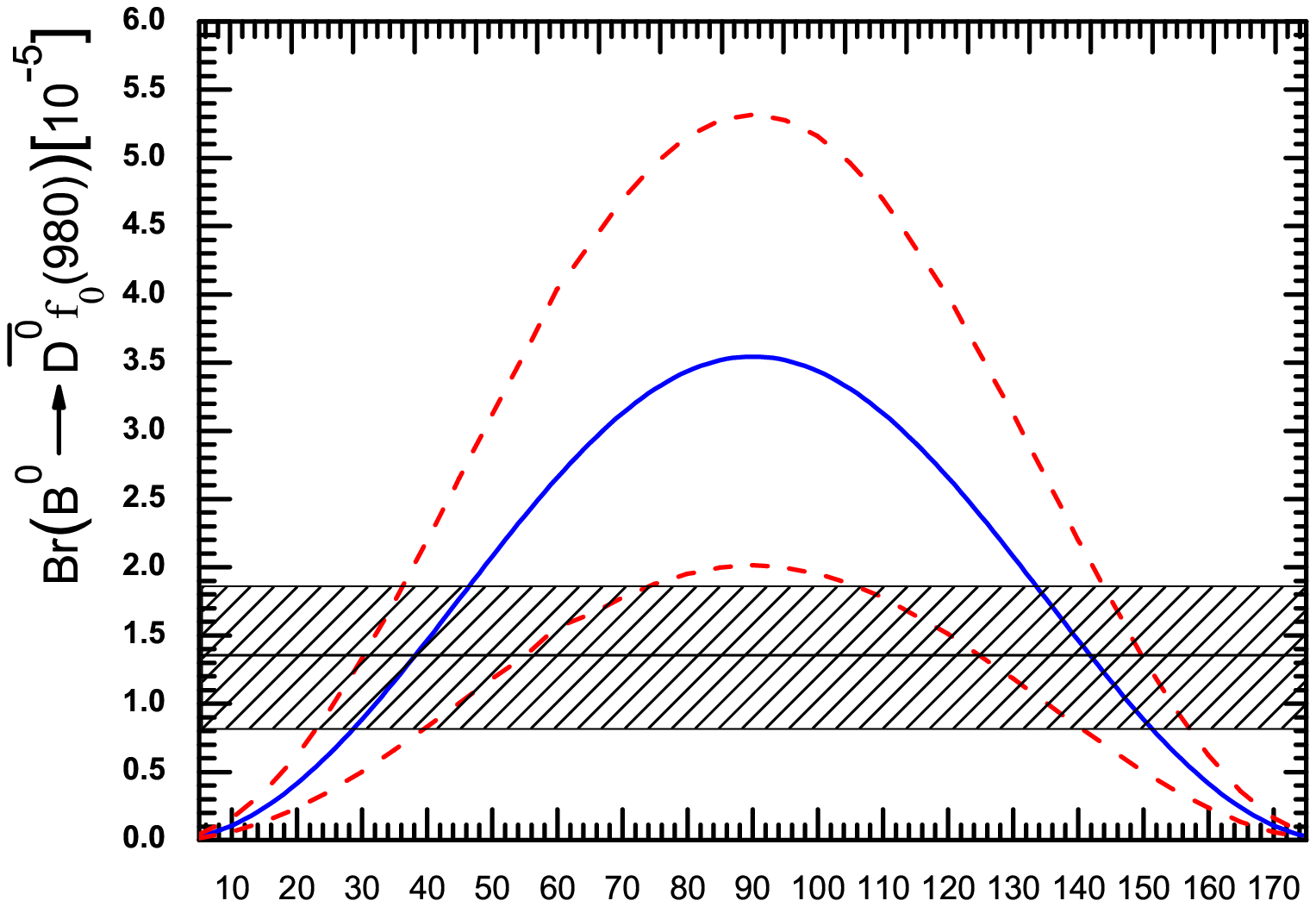} \hspace{-6.5cm} \epsfxsize=15.0cm
\epsffile{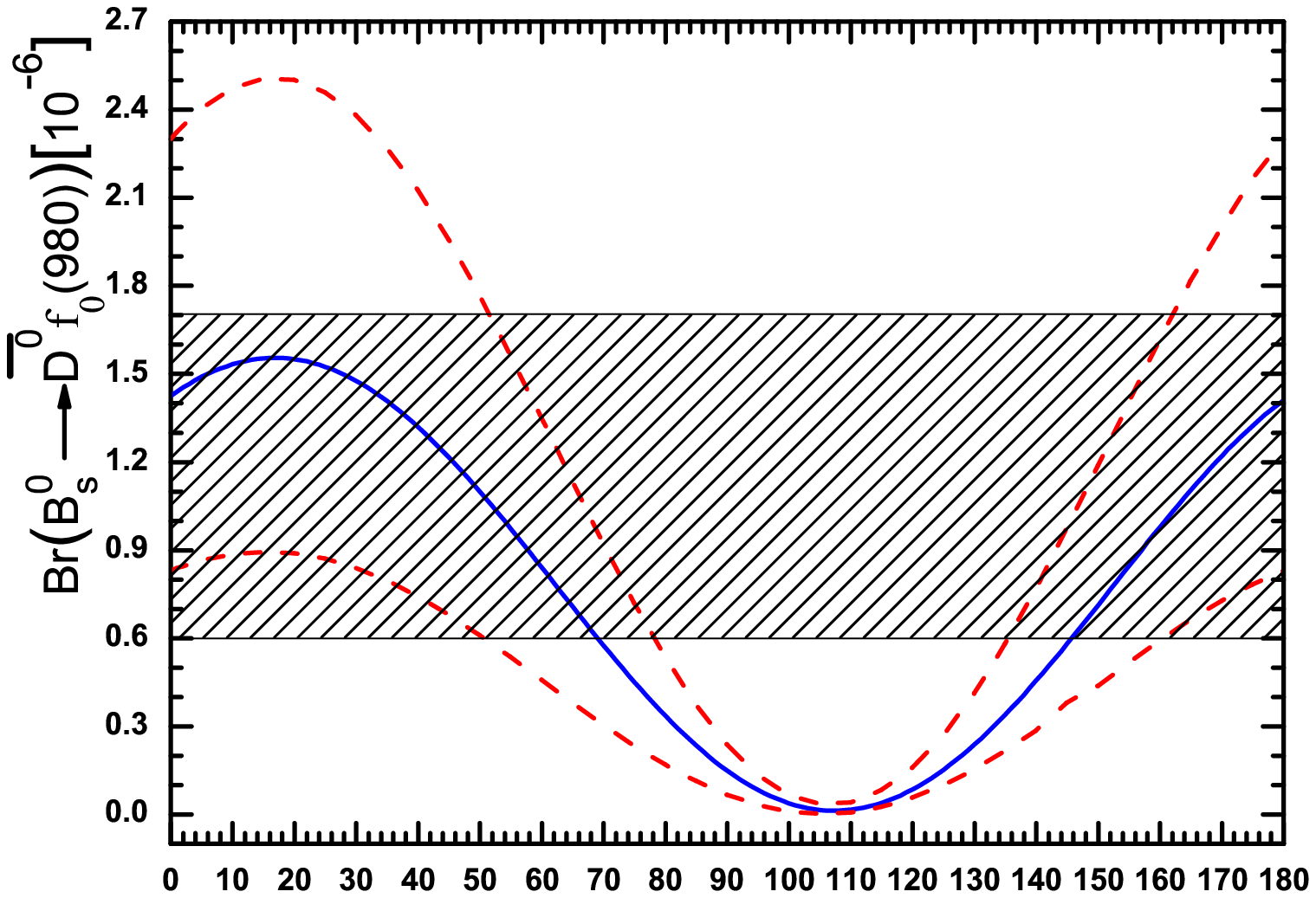}  } \vspace{-4.0cm} \caption{ Dependencies of
the branching ratios $\mathcal{BR}(B^0\to\bar D^0f_0(980))$(left)
and $\mathcal{BR}(B_s\to\bar D^0f_0(980))$ (right) on the mixing
angle $\theta$. In each panel, the solid (blue) curve represents the
central value of the theoretical prediction, and the two dashed
(red) curves correspond to the upper and lower limits. On the left
panel, the shaded band shows the allowed region and the horizontal
bisector  the central value of $\mathcal{BR}(B^0\to\bar
D^0f_0(980))=(1.34\pm0.54)\times10^{-5}$ for data. On the right
panel, for the large uncertainties with the branching ratio, only
the half width band is given, that is to say, the upper edge line
represents the center value of data $\mathcal{BR}(B_s\to\bar
D^0f_0(980))=(1.7\pm1.1)\times10^{-6}$, and the lower edge line
represents the experimental lower limit.    }
%\vspace{-0.5cm}
\label{fig2}
\end{center}
\end{figure}
Combining these two panels, one can find that the allowed mixing
angle lies in the range $135^\circ<\theta<158^\circ$ at the large
angle region. It is not strange that, as mentioned before, the large
mixing angle $\theta\sim146^\circ$ is also obtained in the analysis
of $B_s\to J/\psi f_0(980), J/\psi \sigma$ decays \cite{jingwu}. In
the following we mainly discuss the region with the mixing angle
less than $90^\circ$. For the branching ratio of the decay
$B^0\to\bar D^0f_0(980)$, the experimental value
$(1.34\pm0.54)\times10^{-5}$ with $2.5\sigma$ can give a stronger
constrain on the mixing angle, and in the range of
$29^\circ<\theta<46^\circ$, the central theoretical values agree
well with the data. But if the theoretical uncertainties are
included, the range will become wider. Although the branching ratio
$Br(B_s\to\bar D^0f_0(980))$ with large uncertainty can not give
stringent constrain on the value of the mixing angle, we can get
some hints from the data: If we take the mixing angle
$\theta=0^\circ$, that is, we consider that the scalar meson
$f_0(980)$ is composed entirely of two quark component $s\bar s$,
the corresponding branching ratio is about $1.4\times10^{-6}$, which
is a little lower than the experimental value. If we consider the
small mixing with $q\bar q=(u\bar u+q\bar q)/\sqrt2$, the branching
ratio will get an enhancement for the interference between the two
different kinds amplitudes from the different quark components, the
maximal value for the branching ratio can be obtained at the mixing
angle $\theta=19^\circ$, and arrives at $1.56\times10^{-6}$ (shown
in the right panel of Fig.2). But if we take such small mixing
angle, say about $20^\circ$, it will make the branching ratio of the
decay $B^0\to\bar D^0f_0(980)$ undershoot the shaded band in the
left panel of Fig.2, which represents the experimental allowed
region. While the mixing angle $\theta$ between $f_0(980)$ and
$f_0(500)$ should not be too large, say larger than $70^\circ$. If
so, the predicted branching ratios of both the decays $B_s\to\bar
D^0f_0(980)$ and $B^0\to\bar D^0f_0(980)$ will deviate from the data
even with the large errors taken into account. So we get the
conclusion that the two quark component should be dominant for B
meson decays in dynamic mechanism. Furthermore, the $s\bar s$
component is more important than the $q\bar q$ component. But it is
not in conflict with the dominant four-quark structure in explaining
the mass degeneracy of $f_0(980)$ and $a_0(980)$, and the narrower
decay width of $f_0(980)$ than that of $f_0(500)$. In the following,
we will discuss the mixing angle by considering the ratio of
branching fractions. There are some advantages in considering the
ratio, because one can eliminate the systematic errors on the
experimental side, and avoid the hadronic uncertainties, such as the
decay constants and the Gegenbauer moments of the final states on
the theoretical side. From the data, one can find that the ratio of
these two branching fractions $\mathcal{BR}(B^0\to\bar
D^0f_0(980))/\mathcal{BR}(B_s\to\bar D^0f_0(980))=7.88\pm5.60$.
Unfortunately, here the uncertainty is mainly from the statistical
error in the decay $B_s\to\bar D^0f_0(980)$, so the errors of the
ratio are not much improved compared to those of the branching ratio
of each decay mode. Certainly, here we consider a simple method,
maybe there is a much better approach for the experimentalists to
greatly reduce the errors from this ratio. So we advice to
accurately measure this ratio in experiment, because it is important
to further restrict the mixing angle $\theta$ between $f_0(980)$ and
$f_0(500)(\sigma)$. The ratio can change in a very large range with
the mixing angle taking different values, especially for
$\theta=90^\circ$, the branching ratio of $B_s\to\bar D^0f_0(980)$
is very small and will be exactly equal to zero if the contribution
from $q\bar q=(u\bar u+d\bar d)/\sqrt2$ is turned off, while
$\mathcal{BR}(B^0\to\bar D^0f_0(980))$ arrives its maximal value.
Then it will be meaningless for the ratio, not mentioning the
errors. For the sake of comparison, we give two regions for the
mixing angle shown in Fig.3. If combining these four panels in Fig.2
and Fig.3 together, one will get two further shrunken mixing angle
ranges $22^\circ<\theta<58^\circ$ and $141^\circ<\theta<158^\circ$.
In view of present large uncertainties from data and theory, it will
be difficult to get an unitary value for the mixing angle. But even
if more precise data are available, we still can not get the unitary
value. This argument might be reasonable that there must be some
influence from other components in $f_0(980)$, such as gluon, four
quark component, and $K\bar K$ threshold effect, which we can not
handle at present. Nevertheless, one can not deny that the two quark
component in $f_0(980)$ is dominant in B decay dynamic mechanism,
and the $s\bar s$ component is more important than the $q\bar q$
component.
\begin{figure}[thb]
\begin{center}
\vspace{-0.5cm}
\centerline{\hspace{5.0cm}\epsfxsize=16.0cm \epsffile{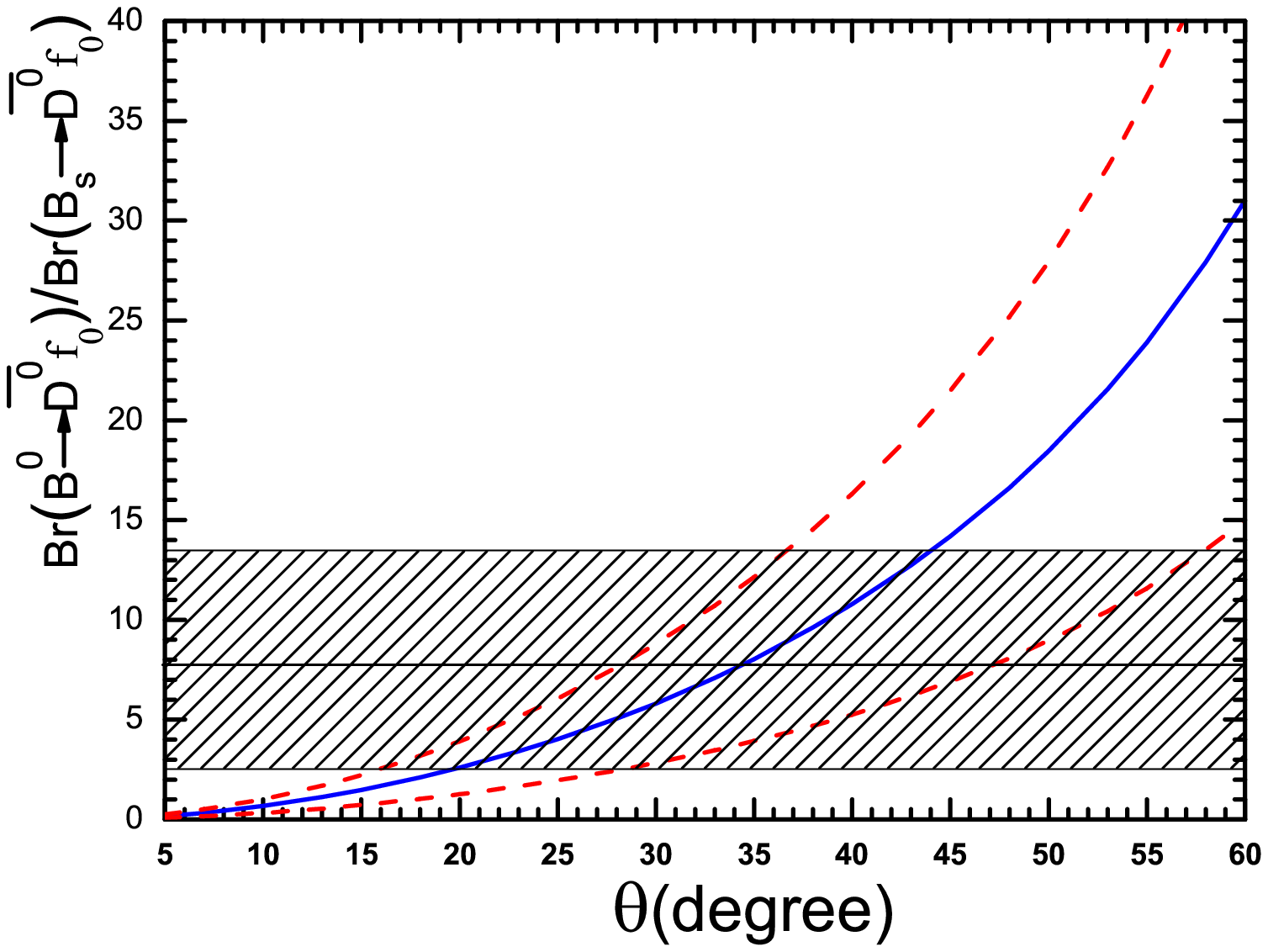} \hspace{-7.5cm}
\epsfxsize=16.0cm \epsffile{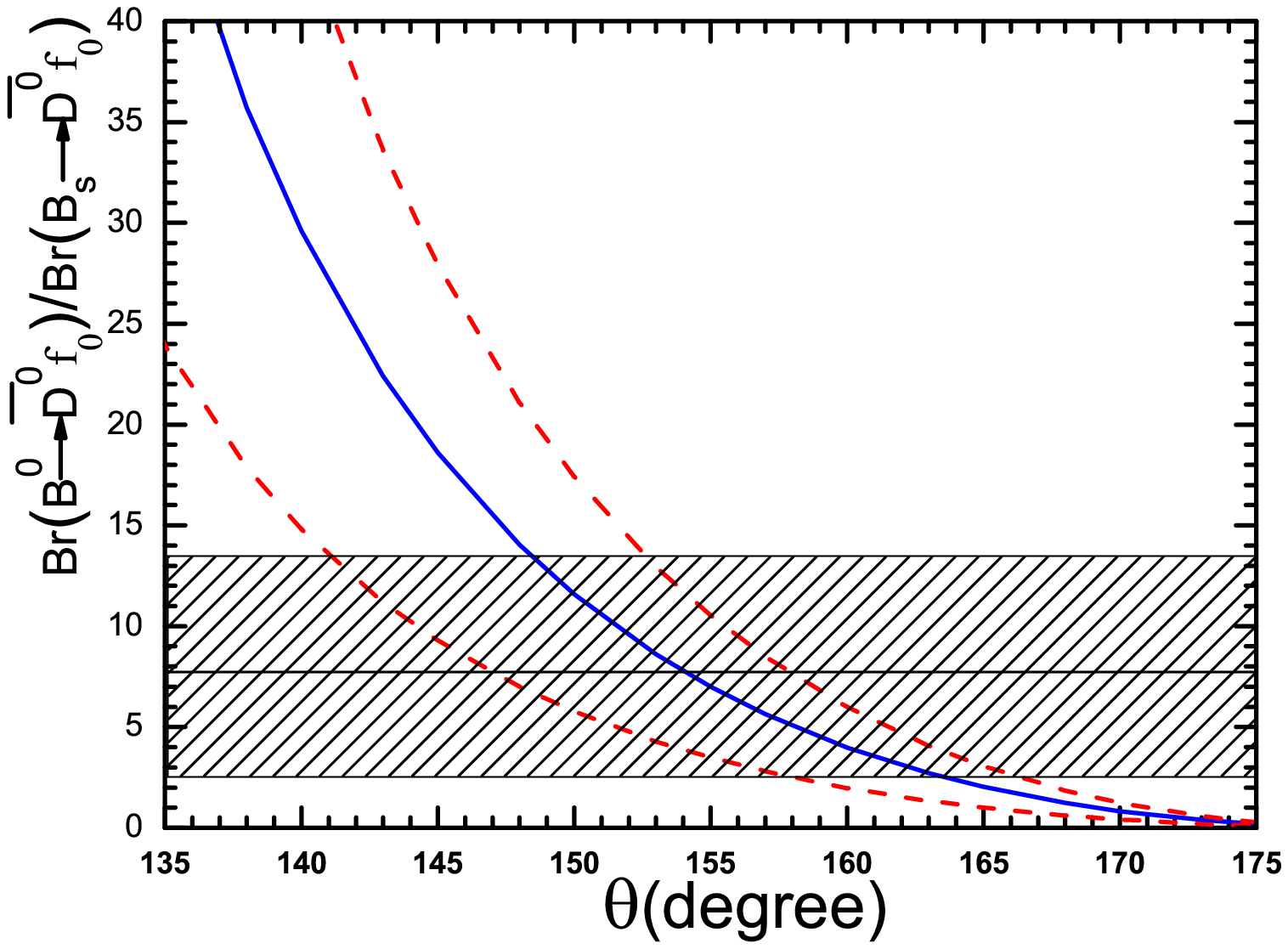}  }
\vspace{-4.0cm}
\caption{ Dependencies of the ratio between $\mathcal{BR}(B^0\to\bar D^0f_0(980))$ and
$\mathcal{BR}(B_s\to\bar D^0f_0(980))$ on the mixing angle $\theta$ at different regions. The shaded band
shows the allowed region and the horizontal bisector the central
value of $\mathcal{BR}(B^0\to\bar D^0f_0(980))/\mathcal{BR}(B_s\to\bar D^0f_0(980))=7.88\pm5.60$ for data.}
%\vspace{-0.5cm}
\label{figs-dep}
\end{center}
\end{figure}

Up to now we still do not analyze the decay $B^0\to \bar D^0
\sigma$, although the data of this channel is available. There are
many uncertainties from the decay constant and the light-cone
distribution amplitudes (LCDAs) of $\sigma$ meson. The authors of
Ref.\cite{CCYscalar} assumed that $\sigma$ has the similar decay
constant and LCDAs as those of $f_0(980)$, while the authors of
Ref.\cite{lirh} just took the same decay constant and LCDAs with
those of $a_0(980)$. These two sets of parameters will generate very
different results: If using the former, one will obtain small
branching ratios which are far below the experimental lower limit in
all the mixing angle region, but the predicted branching ratio will
overlap with the data in some angle values by using the latter,
which can be found in Fig.4. It shows that the decay constant and
LCDAs of $\sigma$ is more close to those of $a_0(980)$, so they
should have the similar quark components and structure. From Fig.4,
we find that there also exist two allowed mixing angle regions
$28^\circ\sim 64^\circ$ and $116^\circ\sim 152^\circ$, where the
former region can overlap mostly with the allowed region
$22^\circ\sim 58^\circ$ obtained from the analysis of $B^0\to \bar
D^0 f_0(980)$ and $B_s\to \bar D^0 f_0(980)$ decays. While the two
large angle regions have less coincidence, it seems that the small
angle region is more favored than the large one.
\begin{figure}[thb]
\begin{center}
\centerline{\hspace{6.0cm} \epsfxsize=18 cm \epsffile{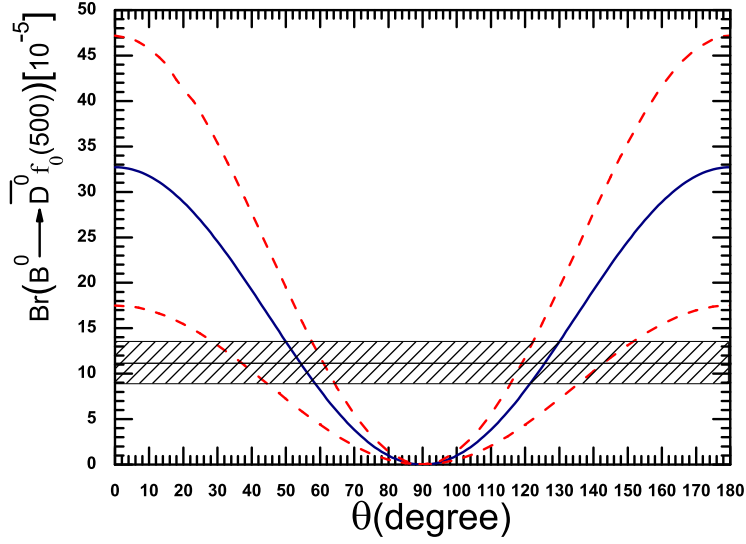}}
\vspace{-4.0cm}
\caption{ Dependence of the branching ratio $\mathcal{BR}(B^0\to\bar D^0f_0(500))$ on the mixing angle $\theta$.
The solid (blue) curve
represents the central value of the theoretical prediction, and the two dashed (red) curves correspond to the upper and lower limits.
The shaded band shows the allowed region and the horizontal bisector the central
value of $\mathcal{BR}(B^0\to\bar D^0f_0(500))=(11.2\pm2.4)\times10^{-5}$ for data.    }
%\vspace{-0.5cm}
\label{fig4}
\end{center}
\end{figure}

\begin{table}
\caption{The CP-averaged branching ratios ($\times10^{-6}$) of $B\to
Df_0(980)$ obtained by taking the mixing angle $\theta=34^\circ$ and
$38^\circ$, respectively. The first uncertainty comes from the
$\omega_b=0.4\pm0.1(0.5\pm0.1)$ for $B(B_s)$ mesons, the second and
the third uncertainties are from the decay constant $\bar
f_{f_0}=0.18\pm0.015$ GeV and the Gegenbauer moment
$B_1=-0.78\pm0.08$ of $f_0(980)$ meson, respectively, and the last
one comes from $C_{D_{(s)}}=0.5(0.4)\pm0.1$ for $D_{(s)}$ meson. }
\begin{center}
\begin{tabular}{c|c|c}
\hline\hline & $34^\circ$ &$38^\circ$  \\
\hline
$\mathcal{BR}(B\to D^0 f_0)[\times10^{-9}]$     &$4.45^{+2.25+0.96+0.71+0.35}_{-1.42-0.85-0.63-0.33}$&$5.39^{+2.72+1.16+0.86+0.43}_{-1.71-1.04-0.77-0.41}$ \\
$\mathcal{BR}(B_{s}\to D^0 f_0)[\times10^{-7}]$ &$1.32^{+1.02+0.30+0.21+0.19}_{-0.60-0.27-0.20-0.17}$&$1.29^{+0.99+0.29+0.21+0.18}_{-0.55-0.27-0.19-0.16}$\\
$\mathcal{BR}(B^+\to D^+f_0)[\times10^{-7}]$    &$1.00^{+0.37+0.16+0.06+0.01}_{-0.26-0.15-0.06-0.01}$&$1.22^{+0.45+0.19+0.08+0.01}_{-0.32-0.18-0.08-0.01}$\\
$\mathcal{BR}(B^+\to D^+_sf_0)[\times10^{-6}]$  &$2.30^{+0.96+0.32+0.11+0.07}_{-0.67-0.30-0.11-0.06}$&$2.97^{+1.20+0.43+0.16+0.07}_{-0.83-0.40-0.15-0.07}$\\
\hline\hline
\end{tabular}\label{tab1}
\end{center}
\end{table}

In order to predict other $B_{(s)}$ charmed decays, the mixing angle
is taken as two values $34^\circ$ and $38^\circ$ (certainly, one can
get similar branching ratios by taking $\theta=142^\circ$ and
$154^\circ$, if they can not be excluded by the future data), one of
which is consistent with $\theta=30^\circ\pm3^\circ$ obtained by
averaging over several processes \cite{ochs}. Then the branching
ratios of these CKM suppressed decays $B^0\to D^0 f_0(980), B_{s}\to
D^0 f_0(980), B^+\to D^+f_0(980)$ and $B^+\to D^+_sf_0(980)$  are
listed in Table \ref{tab1}. When the pseudoscalar meson $D_{(s)}$ is
replaced by the vector meson $D^*_{(s)}$ in our considering decays,
and the branching ratios of the corresponding channels are listed in
Table \ref{tab2}. From our calculations, we find that the branching
ratios of the $B_s$ decays are not very sensitive to the mixing
angle $\theta$, especially for $\mathcal{BR}(B_{s}\to D^0 f_0)$, its
value changes in the range of $(1.2\sim1.8)\times10^{-7}$ when the
mixing angle varies from $0^\circ$ to $180^\circ$. The reason is as
follows: The amplitude from $s\bar s$ component has a large
imaginary part and a small real part. It is contrary for the
amplitude from $q\bar q=(u\bar u+d\bar d)/\sqrt2$ component, where
the real part is about one order larger than the imaginary part.
When the real and imaginary parts from the $s\bar s$ and $q\bar
q=(u\bar u+d\bar d)/\sqrt2$ amplitudes mixing through
Eq.(\ref{mix}), respectively, the former (later) is dominated by the
sine (cosine) law, but the later is stronger than the former, so
these two kinds of contrary change trends make the total amplitude
changes in a much milder cosine curve. The branching ratios of all
the $B$ decay modes are dependent on the mixing angle via
$\sin\theta$ (maybe with an initial phase), just like the left panel
in Fig.\ref{fig2}, while those of $B_s$ decay modes are dependent on
the mixing angle via $\cos\theta$ with a initial phase, just like
the right panel in Fig.\ref{fig2}.

\begin{table}
\caption{Same as Table.\ref{tab1} except for the decays $B\to \bar D^*(D^*)f_0(980)$. }
\begin{center}
\begin{tabular}{c|c|c}
\hline\hline & $34^\circ$ &$38^\circ$  \\
\hline
$\mathcal{BR}(B\to \bar D^{*0} f_0)[\times10^{-6}]$     &$7.40^{+2.33+1.32+2.32+0.75}_{-1.84-1.26-1.78-0.73}$&$8.97^{+2.83+1.60+2.82+0.91}_{-2.23-1.52-2.16-0.89}$ \\
$\mathcal{BR}(B_{s}\to \bar D^{*0} f_0)[\times10^{-6}]$ &$1.63^{+0.72+0.31+0.48+0.20}_{-0.50-0.29-0.38-0.17}$&$1.43^{+0.62+0.27+0.42+0.17}_{-0.44-0.25-0.33-0.15}$\\
$\mathcal{BR}(B\to D^{*0} f_0)[\times10^{-9}]$     &$6.48^{+3.57+1.37+0.64+0.33}_{-2.24-1.23-0.56-0.31}$&$7.86^{+4.33+1.66+0.78+0.40}_{-2.72-1.49-0.68-0.37}$ \\
$\mathcal{BR}(B_{s}\to D^{*0} f_0)[\times10^{-7}]$ &$2.06^{+1.79+0.46+0.20+0.20}_{-0.98-0.41-0.18-0.17}$&$1.94^{+1.63+0.44+0.19+0.19}_{-0.90-0.39-0.17-0.16}$\\
$\mathcal{BR}(B^+\to D^{*+}f_0)[\times10^{-7}]$    &$2.07^{+0.69+0.38+0.16+0.02}_{-0.49-0.34-0.15-0.02}$&$2.51^{+0.84+0.46+0.19+0.02}_{-0.60-0.42-0.19-0.03}$\\
$\mathcal{BR}(B^+\to D^{*+}_sf_0)[\times10^{-6}]$  &$5.00^{+1.68+0.94+0.37+0.07}_{-1.21-0.88-0.39-0.06}$&$6.10^{+2.04+1.14+0.45+0.08}_{-1.47-1.06-0.47-0.10}$\\
\hline\hline
\end{tabular}\label{tab2}
\end{center}
\end{table}
%%%%%%%%%%%%%%%%%%%%%%%%%%%%%%%%%%%%%%%%
%=================================================================
%===========================================================================
%                 Conclusion
%============================================================================

\section{Conclusion}\label{summary}

In this paper, first we analyze the decays $B\to \bar D^{0}
f_0(980)$ and $B_{s}\to \bar D^{0} f_0(980)$ carefully in the PQCD
factorization approach and find two possible regions for the mixing
angle $\theta$, one is centered at $34^\circ\sim38^\circ$ and the
other is near $142^\circ\sim154^\circ$. If the data of the decay
$B^0\to \bar D^{0} \sigma$ is also included, we find that the small
angle region is more favored. Our analyses support that the two
quark component in $f_0(980)$ is dominant in B decay dynamic
mechanism, and the $s\bar s$ component is more important than the
$q\bar q$ component. Certainly other components, such as gluon, four
quark component, and $K\bar K$ threshold effect may also give some
more or less influences. It is noticed that our picture is not in
conflict with the popular explanation of dominant four-quark
component in $f_0(980)$. Then we predict the branching ratios of
other $B_{(s)}\to D_{(s)} f_0(980),D^{*}_{(s)} f_0(980)$ decay
channels by fixing $\theta=34^\circ$ and $38^\circ$, respectively
and find that the branching ratios of $B_s$ decay modes are less
sensitive to the mixing angle compared with those of $B$ decay
modes. Especially, for the decay $B_{s}\to D^0 f_0$, its branching
ratio changes in a small region between $(1.2\sim1.8)\times10^{-7}$
with the mixing angle $\theta$ running from $0^\circ$ to
$180^\circ$.
%%%%%%%%%%%%%%%%%%%%%%%%%%%%%%%%%%%%%%%%%%%%%%%%%%%%%%%%%%%%%%%%%%%%%%%%%%%%%%%
\section*{Acknowledgment}
This work is partly supported by the National Natural Science
Foundation of China under Grant No. 11347030, by the Program of
Science and Technology Innovation Talents in Universities of Henan
Province 14HASTIT037. Z.Q. Zhang is grateful to Hai-Yang Cheng and
Hsiang-nan Li for carefully reading the manuscript and very useful
suggestions, Wen-Fei Wang, Henry T. Wong, Tzu-Chiang Yuan for
helpful discussions. He also thanks the Institute of Physics,
Academia Sinica for their hospitalities during his visit when part
of this work was done.
\appendix
\section{Decay amplitudes}
For the CKM suppressed decays, for example, $B\to D^0f_0(980)$,
their Feynman diagrams to leading order will be different from
Fig.1, especially for the (non-)factorizable annihilation diagrams,
where the positions of $D$ and $f_0(980)$ are exchanged compared
with those of $B\to \bar D^0f_0(980)$ decay. But the factorizable
emission diagrams are the same with each other, so
$\mathcal{F}^{D}_{B\to f_0}=\mathcal{F}^{\bar D}_{B\to f_0}$. Here
we also list other amplitudes of these CKM suppressed decays: \be
\mathcal{M}^{D}_{B\to f_0}&=&32\pi C_f m_B^4/\sqrt{2N_C}\int_0^1 d
x_{1} dx_{2} dx_{3}\, \int_{0}^{\infty} b_1 db_1 b_3
db_3\,\phi_B(x_1,b_1)\phi_D(x_3,b_3)\non &&
\times\left\{\left[(x_3-x_1)\phi_{f_0}(x_2)-r_{f_0}x_2(\phi^s_{f_0}(x_2)-\phi^t_{f_0}(x_2))\right]
\right.\non &&\left.\times
E_{en}(t_d)h^{d}_{en}(x_1,x_2(1-r_D^2),x_3,b_1,b_3)+E_{en}(t_c)h^{c}_{en}(x_1,x_2(1-r_D^2),x_2,b_1,b_3)
\right.\non && \left. \times\left[(x_1-x_2+x_3-1)\phi_{f_0}(x_2)
+r_{f_0}x_2(\phi^s_{f_0}(x_2)+\phi^t_{f_0}(x_2))\right]\right\},\\
\mathcal{M}^{f_0}_{ann}&=&32\pi C_f m_B^4/\sqrt{2N_C}\int_0^1 d
x_{1} dx_{2} dx_{3}\,\int_{0}^{\infty} b_1 db_1 b_3 db_3\,
\phi_B(x_1,b_1)\phi_D(x_3,b_3)\non &&
\times\left\{E_{an}(t_{e})h^e_{an}(x_1,x_2,x_3,b_1,b_3)\left[(1-r_b-x_2)\phi_{f_0}(x_2)
\right.\right.\non && \left.\left.
+r_Dr_{f_0}((2-4r_b-x_2-x_3)\phi^s_{f_0}(x_2)-(x_2-x_3)\phi^t_{f_0}(x_2))
\right]\right.\non &&\left.
+E_{an}(t_{f})h^f_{an}(x_1,x_2,x_3,b_1,b_2)\left[x_3 \phi_{f_0}(x_2)
\right.\right.\non &&\left.\left.
+r_Dr_{f_{0}}((x_2+x_3)\phi^s_{f_{0}}(x_2)+(x_3-x_2)\phi^t_{f_{0}}(x_2))\right]\right\},\\
\mathcal{F}^{f_0}_{ann}&=&8\pi C_f f_{B}m_B^4\int_0^1 d x_{2}
dx_{3}\, \int_{0}^{\infty} b_2 db_2 b_2 db_2\,
\phi_D(x_3,b_3)\left\{\left[(r_D^2-1)x_3\phi_{f_{0}}(x_2)\right.\right.\non
&& \left.\left.-2r_{f_0}r_D(1-r_D^2+x_3)\phi^s_{f_{0}}(x_2)\right]
E_{af}(t'_g)h_{af}(x_3,(1-x_2)(1-r_D^2),b_3,b_2)\right.\non &&\left.
+E_{af}(t'_h)h_{af}(x_2,x_3(1-r_D^2),b_2,b_3)\left[(x_2-2r_Dr_c)\phi_{f_{0}}(x_2)
\right.\right.\non &&\left.\left.
+2r_Dr_{f_{0}}((x_2+1)\phi^s_{f_{0}}(x_2)+(x_2-1)\phi^t_{f_{0}}(x_2)\right]\right\}.
\en Here we do not show the amplitudes of the decays $B_{(s)}\to
\bar D^{*}(D^{*})f_0(980)$, because one can obtain them from those
of the decays $B_{(s)}\to \bar D(D) f_0(980)$ by the substitutions
$m_D\rightarrow m_{D^*}, f_D\rightarrow f_{D^*}, \phi_D\rightarrow
\phi_{D^*}$, where the terms including $r_D^2, r_Dr_{f_0}$ and
$r_Dr_c$ were neglected. It is similar for the decays involving
$D^{*}_s$ meson.
\section{Hard scales }
\be
t_a&=&\max(\sqrt{x_2(1-r^2_D)}m_B,1/b_1,1/b_2),\\
t_b&=&\max(\sqrt{x_1(1-r^2_D)}m_B,1/b_1,1/b_2),\\
t_{c,d}&=&\max(\sqrt{x_1x_2(1-r_D^2)}m_B,\sqrt{|A^2_{c,d}|}m_B,1/b_1,1/b_3),\\
t_{e,f}&=&\max(\sqrt{x_2x_3(1-r_D^2)}m_B,\sqrt{|L^2_{e,f}|},m_B,1/b_1,1/b_3),\\
t_{g}&=&\max(\sqrt{(1-x_2)(1-r^2_D)}m_B,1/b_2,1/b_3), \\
t_{h}&=&t'_g=\max(\sqrt{x_3(1-r^2_D)}m_B,1/b_2,1/b_3),\\
t'_{h}&=&\max(\sqrt{x_2(1-r^2_D)}m_B,1/b_2,1/b_3). \en And the
$S_j(t) (j=B,D,f_0)$ functions in Sudakov form factors in
Eq.(\ref{suda1}), Eq.(\ref{suda2}), Eq.(\ref{suda3}) and
Eq.(\ref{suda4}) are listed as \be
S_B(t)&=&s(x_1\frac{m_B}{\sqrt2},b_1)+2\int^t_{1/b_1}\frac{d\bar\mu}{\bar\mu}\gamma_q(\alpha_s(\bar\mu)),\\
S_{D}(t)&=&s(x_3\frac{m_B}{\sqrt2},b_3)+2\int^t_{1/b_3}\frac{d\bar\mu}{\bar\mu}\gamma_q(\alpha_s(\bar\mu)),\\
S_{f_0}(t)&=&s(x_2\frac{m_B}{\sqrt2},b_2)+s((1-x_2)\frac{m_B}{\sqrt2},b_2)+2\int^t_{1/b_2}\frac{d\bar\mu}{\bar\mu}\gamma_q(\alpha_s(\bar\mu)),
\en where the quark anomalous dimension is $\gamma_q=-\alpha_s/\pi$,
and the expression of the $s(Q,b)$ in one-loop running coupling
coupling constant is used \be
s(Q,b)&=&\frac{A^{(1)}}{2\beta_1}\hat{q}\ln(\frac{\hat{q}}{\hat{b}})-\frac{A^{(1)}}{2\beta_1}(\hat{q}-\hat{b})
+\frac{A^{(2)}}{4\beta^2_1}(\frac{\hat{q}}{\hat{b}}-1)\non
&&-\left[\frac{A^{(2)}}{4\beta^2_1}-\frac{A^{(1)}}{4\beta_1}\ln(\frac{e^{2\gamma_E-1}}{2})\right]
\ln(\frac{\hat{q}}{\hat{b}}), \en with the variables are defined by
$\hat{q}=\ln[Q/(\sqrt2\Lambda)], \hat{q}=\ln[1/(b\Lambda)]$ and the
coefficients $A^{(1,2)}$ and $\beta_{1}$ are \be
\beta_1&=&\frac{33-2n_f}{12},A^{(1)}=\frac{4}{3},\\
A^{(2)}&=&\frac{67}{9}-\frac{\pi^2}{3}
-\frac{10}{27}n_f+\frac{8}{3}\beta_1\ln(\frac{1}{2}e^{\gamma_E}),
\en where $n_f$ is the number of the quark flavors and $\gamma_E$ the
Euler constant.
%%%%%%%%%%%%%%%%%%%%%%%%%%%%%%%%%%%%%%%%%%%%%%%%%%%%%%%%%%%%%%%%%%%%%%%%
%                               references
%%%%%%%%%%%%%%%%%%%%%%%%%%%%%%%%%%%%%%%%%%%%%%%%%%%%%%%%%%%%%%%%%%%%%%%%


\begin{thebibliography}{99}
\bibitem{jaffe}
R.L. Jaffe, Phys.Rev.D {\bf15}, 267 (1977); ibid.281 (1977).
\bibitem{minkowski}
P. Minkowski and W. Ochs, Eur.Phys.J.C. {\bf9}, 283 (1999).
\bibitem{gokalp}
A. Gokalp, Y. Sarac, O.Yilmaz, Phys.Lett.B{\bf609}, 291 (2005).
\bibitem{jingwu}
J.W. Li, D.S. Du, and C.D. Lu, \epjc {\bf72}, 2229 (2012).
\bibitem{ochs}
W.Ochs, J.Phys.G {\bf40}, 043001 (2013).
\bibitem{lmzhang}
S. Stone, L. Zhang, Phys.Rev.Lett. {\bf111}, 062001 (2013).
\bibitem{feldmann1}
T.Feldmann, P.Kroll, and B. Stech, \prd{\bf58},114006 (1998).
\bibitem{feldmann2}
T.Feldmann, P.Kroll, and B. Stech, \plb{\bf449}, 339 (1999).
\bibitem{hnli1}
Y.D. Tsai, H.n.Li, Q. Zhao, \prd {\bf85}, 034002 (2012).
\bibitem{lhcb1}
LHCb Collaboration, R.Aaij{\it et al.}, Phys.Rev.D {\bf92}, 032002
(2015).
\bibitem{lhcb2}
LHCb Collaboration, R.Aaij{\it et al.}, JHEP {\bf08}, 005 (2015).
\bibitem{li}
R.H. Li, C.D. Lu, and H. Zou, \prd {\bf79}, 014013 (2009).
\bibitem{zou}
H. Zou, R.H. Li, X.X. Wang and C.D. Lu, J.Phys.G {\bf37}, 015002
(2010).
\bibitem{zhang1}
Z.Q. Zhang, J.Phys.G {\bf36}, 125004 (2009).
\bibitem{zhang2}
Z.Q. Zhang, \prd {\bf87}, 074030 (2013).
\bibitem{kim}
C.S. Kim, R.H. Li and W. Wang, \prd{\bf88}, 034003 (2013).
\bibitem{CCYscalar} H. Y. Cheng, K. C. Yang, \prd {\bf73}, 014017 (2006).
\bibitem{lirh}
R.H. Li, C.D. Lu, W. Wang, and X.X. Wang,\prd{\bf79}, 014013 (2009).
\bibitem{cdlu}
C.D. Lu, M.Z. Yang, Eur. Phys. J. C{\bf28}, 515 (2003).
\bibitem{kurimoto}
T. Kurimoto, H.-n. Li and A.I. Sanda, \prd {\bf67}, 054028 (2003).
\bibitem{pdg14}
K.A.Olive {\it et al.}[Particle Data Group Collaboration],
Chin.Phys.C {\bf38}, 090001 (2014).
\bibitem{wanggl}
D.Becirevic, Ph.Boucaud, J.P. Leroy, V.Lubicz, G.Martinelli,
F.Mescia, and F.Rapuano, \prd{\bf60},074501 (1999); M.A. Ivanov,
J.G. Korner, P.Santorelli, \prd{\bf73}, 054024 (2006); G.L. Wang,
\plb{\bf633}, 492 (2006).
\bibitem{defa}
F.De Fazio and M.R. Pennington, Phys.Lett.B {\bf521}, 15 (2001).
\bibitem{hk}
Hai-Yang Cheng and Kwei-Chou Yang, \prd {\bf76}, 114020 (2007).
\end{thebibliography}
\end{document}